\documentclass[conference]{IEEEtran}
\IEEEoverridecommandlockouts
\usepackage{cite}
\usepackage{amsmath,amssymb,amsfonts}
\usepackage{algorithmic}
\usepackage{graphicx}
\usepackage{textcomp}
\usepackage{xcolor}
\usepackage{listings}
\usepackage{soul}
\def\BibTeX{{\rm B\kern-.05em{\sc i\kern-.025em b}\kern-.08em
    T\kern-.1667em\lower.7ex\hbox{E}\kern-.125emX}}
\usepackage{hyperref}
\hypersetup{
	citecolor = blue,
	linkcolor = blue,
	urlcolor = blue,
	colorlinks = true,
	linkbordercolor = {white}
}

\colorlet{punct}{red!60!black}
\definecolor{background}{HTML}{EEEEEE}
\definecolor{delim}{RGB}{20,105,176}
\colorlet{numb}{magenta!60!black}

\lstdefinelanguage{json}{
	basicstyle=\normalfont\ttfamily,
}

\begin{document}

\title{\textit{CoronaZ}: another distributed systems project
\\{\Large Simulating a contact tracing application in a scalable environment}
}

\author{
	
	\IEEEauthorblockN{Stefan Ciprian Voinea}
	\IEEEauthorblockA{Università degli Studi di Padova}
	\textit{stefanciprian.voinea@studenti.unipd.it}
	\and
	\IEEEauthorblockN{Stefan Vladov}
	\IEEEauthorblockA{Technical University of Munich}
	\textit{stefan.vladov@tum.de}
	\and
	\IEEEauthorblockN{Fabian Rensing}
	\IEEEauthorblockA{Paderborn University}
	\textit{fabian.rensing@helsinki.fi}
}

\maketitle
\thispagestyle{plain}
\pagestyle{plain}



\section{Introduction}\label{sec:introduction}

	This brief paper describes \textit{CoronaZ}, a project for the Distributed Systems course at the \textit{University of Helsinki}.
	
	All the code of the project is publicly available on GitHub repository\cite{coronaz_repo}.
	
	This project simulates a contact tracing application where each node represents a person (or a unique device attached to someone) that send signals to each other when in range and communicate the data collected to a server using the \textit{publish-subscribe} pattern.
	The server, called \textit{broker}, can then be polled by a node called \textit{consumer} that will send the data to a database.
	A front-end application then requests this data and displays the movement and the latest updates via the browser.
	
	The idea came from simulating this kind of movements with Arduino boards capable of communicating between themselves using the \textit{nrf24l01} and to the broker with \textit{esp8266}.
	Unfortunately this was not possible given the relatively strict amount of time that each of the students involved could dedicate to the project and the waiting time to get the necessary hardware.

\section{Technological choices}\label{sec:technological_choices}

	In this section we describe and explain why we have decided to use certain technologies rather than others.
	
	\begin{itemize}
		\item \textit{Programming language}: we chose Python because it is a fast programming language for creating prototypes. Also all group members are fluent with Python;
		\item \textit{MQTT Broker}: we chose \textit{Apache Kafka} as broker for our project since it is one of the most used brokers in the market and it is has a large community that supports the project. Apache Kafka also handles well scaling and integration with other systems;
		\item \textit{Containers}: we chose to use \textit{Docker} and \textit{docker-compose} given the easiness of constructing and spawning nodes. As explained in \ref{sec:architecture}, we have a \texttt{docker-compose.yml} that handles that base components such as the broker and the database;
		\item \textit{Database}: we chose \textit{MongoDB} for its high scalability and ease of use. Additionally, its dynamic schemas would allow us to be very flexible with our data model;
		\item \textit{Back-end}: we chose \textit{NodeJS} and \textit{Express}, as it is a popular back-end for web applications, and some group members had prior experience
		\item \textit{Front-end}: we chose \textit{React} for its fast development speed and familiarity of some group members with it.
	\end{itemize}

	\subsection{System requirements}
	
		The system requirements for this project are simple:
		\begin{itemize}
			\item \textit{Docker} and \textit{docker-compose};
			\item Ubuntu or another Linux system with \texttt{jq} installed for starting the project using the \texttt{init-project.sh} bash script (partly working with \texttt{git bash} on Windows).
		\end{itemize}

\section{Architecture}\label{sec:architecture}
	
	We can divide our project in five major areas: \textit{nodes}, \textit{broker}, \textit{DB consumer}, \textit{database}, \textit{back-end} and \textit{front-end}.

	\begin{figure}[htbp]
		\centerline{\includegraphics[width=\linewidth]{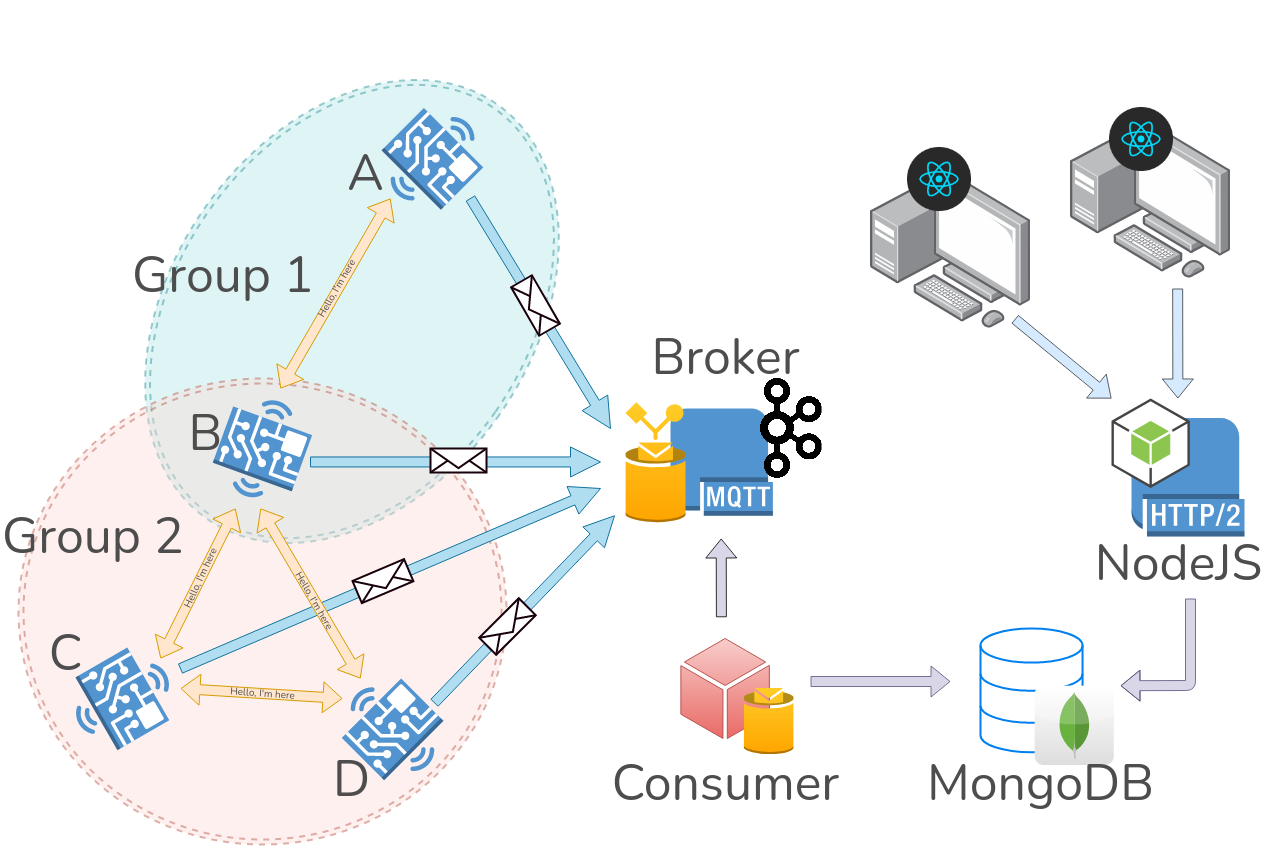}}
		\caption{Architecture of the \textit{CoronaZ} project.}
		\label{fig:architecture}
	\end{figure}
	
	\subsection{Nodes}
	
		This can be considered a single component since each node can be spawned separately from one another.
		
		When each node spawns in the map it is placed on a random position.
		Every second the node will ``\textit{move}'' in a random direction, broadcast a message telling its position and \texttt{UUID} (\textit{Universally Unique Identifier}) and listen to the ones that are in broadcast by other nodes.
		
		Nodes introduced into the simulation can be \textit{infected} or \textit{non-infected} (or \textit{safe}).
		When a node is infected it will stay in his last position and will not move for a certain amount of time as specified in the parameter \texttt{infection\_cooldown}, which defines the number of seconds that the node will stay in place.
		After this amount of time, the node will have an ``\textit{immunity period}'' in which it will be able to move again and, if it gets in contact with an infected node, it will not get infected.
		This time is the same as the \texttt{infection\_cooldown} parameter.
		
		In a gamification perspective, the nodes in our system are also called \textit{humans} and \textit{zombies}. 
		Humans are the nodes considered \textit{safe} while the \textit{zombies} are the nodes that have been \textit{infected}.
		
		In our simulation the nodes can all connect to each other since they are all in the same network, as explained in \ref{sec:network}, and each of them can hear the data sent in broadcast by the others.
		In a more realistic situation nodes would be only capable of hearing the signals of nodes nearby them, like in Fig.\ref{fig:architecture}~where node \textit{A} can communicate only with node \textit{B} and nodes \textit{B}, \textit{C} and \textit{D} are nearby each other enough for them to hear their signal.

		Here is an example of a message that is sent in broadcast from a node:
		\begin{lstlisting}[language=json]
{  
   "uuid":"ff0a1bda-34b9-11eb-b339 ... ",
   "position":[1, 5],
   "infected":false,
   "timestamp":"2020-12-02 16:19 ... ",
   "alive":true
}  
		\end{lstlisting}
		
		After the node exceeds its lifetime in seconds, before exiting will send a last message where the value of \texttt{alive} will be \texttt{false}.
		This signals the front-end that this node will not have to be displayed anymore.
		
		Each node has a unique \textit{UUID} that is generated in Python, using the \texttt{uuid}\cite{uuid}.
		This is created via a combination of \texttt{MAC} address and the \texttt{IP} address of the machine the script runs on and the timestamp on when the process starts.

		The nodes images are built via a \textit{Dockerfile} that uses the small Linux distribution Alpine to run the Python scripts.
		The nodes' python scripts are all loaded in the Docker image and the \texttt{main.py} file is executed as \texttt{CMD} when the container starts.
		This configuration allows to run containers with arguments as well, thus deciding if the node will start as an infected node or as a safe node.
	
		The work of the nodes is split along four \textit{threads}:
		\begin{itemize}
			\item main thread that starts all the other threads and controls them (main logic of the program);
			\item broadcasting thread in which the commands for broadcasting the message are executed;
			\item listening thread that waits for incoming messages;
			\item Kafka server connection thread that checks if the connection with the Kafka server is still up.
		\end{itemize}
		
		For connectivity the nodes use the \texttt{socket library} and will send UDP messages in broadcast to the unassigned port \texttt{4711} via a random port chosen by the library.
		
		The sender's IP address and the port that the message has been sent from can be seen in the logs printed on screen.
		Setting the debugging at \texttt{INFO} level helps seeing the messages that each node send with one another and that are sent to Kafka.
		
		The messages to be sent to the Kafka server are collected via the \texttt{get\_next\_broadcast\_message} method and sent using the \texttt{send\_message} method.
	
	
	\subsection{MQTT broker}	
	
		Apache Kafka is composed by the Kafka container and the Zookeeper container.
		Kafka is an important part in the project since it is the \textit{broker} in the \textit{publish-subscribe} pattern.
		
		Each node, after it has collected the IDs of the other nearby it, will send the list ot these IDs, with other information, to the broker.
		The \textit{topic}, in our code, used by the nodes is ``\textit{coronaz}''.
		This will contain all the information send by the nodes to the broker, which will later forward them to the consumer when it asks for them.
		
		Here is an example of a message that is sent to the broker:
		\begin{lstlisting}[language=json]
{
  "uuid": "5603b252-36de-11eb ... ",
  "position": [
      72,
      33
    ],
  "infected": false,
  "timestamp": "2020-12-05 09:43 ... ",
  "alive": true,
  "contacts": [
    {
      "uuid": "563eafe6-36de-11eb ... ",
      "timestamp": "2020-12-05 09:43 ... "
    },
    {
      "uuid": "56fd5d12-36de-11eb ... ",
      "timestamp": "2020-12-05 09:43 ... "
    },
    {
      "uuid": "567a64c7-36de-11eb ... ",
      "timestamp": "2020-12-05 09:43 ... "
    },
    {
      "uuid": "56b292fc-36de-11eb ... ",
      "timestamp": "2020-12-05 09:43 ... "
    },
    {
      "uuid": "575030c3-36de-11eb ... ",
      "timestamp": "2020-12-05 09:43 ... "
    },
    {
      "uuid": "563eafe6-36de-11eb ... ",
      "timestamp": "2020-12-05 09:43 ... "
    }
  ]
}
		\end{lstlisting}
		
		If a message is received from the same node twice then, the receiving node will discard the first message that arrived.
	
	\subsection{DB consumer}
	
		The DB consumer can be considered as a single entity since it is independent both from Kafka and from Mongo.
		The consumer is subscribed to the Kafka topic that contains the new messages from the nodes, in our case ``\textit{coronaz}''.
		When a new message arrives, the consumer gets it and, every ten messages (or when a node dies), it aggregates them in a \texttt{json} that will be sent to the MongoDB database.
	
	\subsection{Database}
	
		As the other components, the database runs in a Docker container.
		
		It accepts and stores incoming data from the Consumer.
		The database is accessed by the back-end, which will forward the data to the front-end, refreshed every second in order to always display the latest status.
	
		\begin{figure}[htbp]
			\centerline{\includegraphics[width=\linewidth]{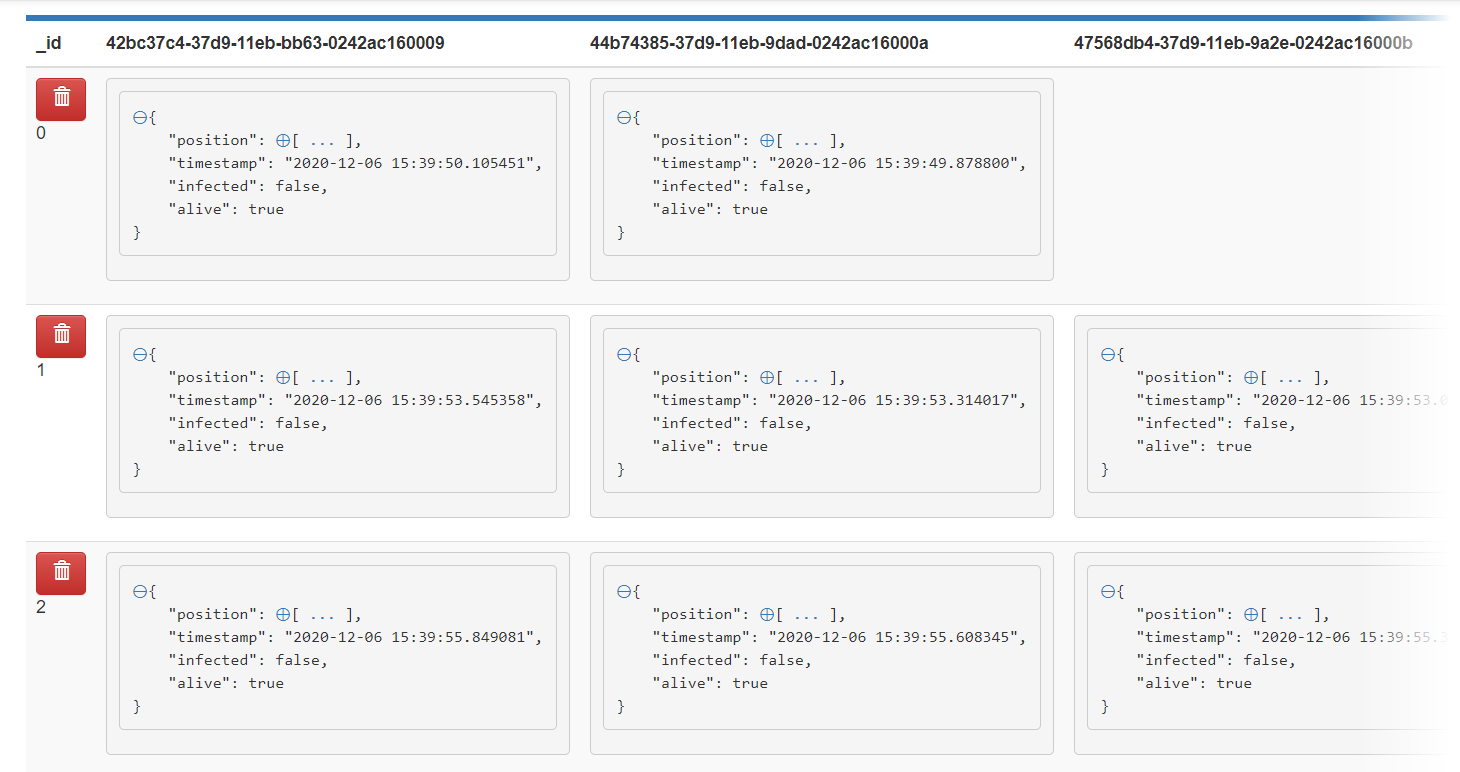}}
			\caption{CoronaZ database structure.}
			\label{fig:database}
		\end{figure}
	
		In the database, each document is mapped with the \texttt{UUID} of the nodes as a key and the value represents the latest update of the node(from the last message).
	
	\subsection{Back-end}
	
		For the back-end of the project, \textit{NodeJS} provides a single REST endpoint for the front-end. 
		This is \textit{GET /data}, which connects to the MongoDB database and queries all the documents containing the state of the nodes.
	
	\subsection{Front-end}

		The front-end of the project, built with \textit{React} and \textit{MaterialUI}, shows the evolution of the system and the simulations.
		
		A slider allows the user to select one of the queried states from the database, which will be then visualized on the map underneath. 
		Moreover, this will also update the statistics on the top of the page.
		Realism mode switches between using circles or icons.
		
		\begin{figure}[htbp]
			\centerline{\includegraphics[width=\linewidth]{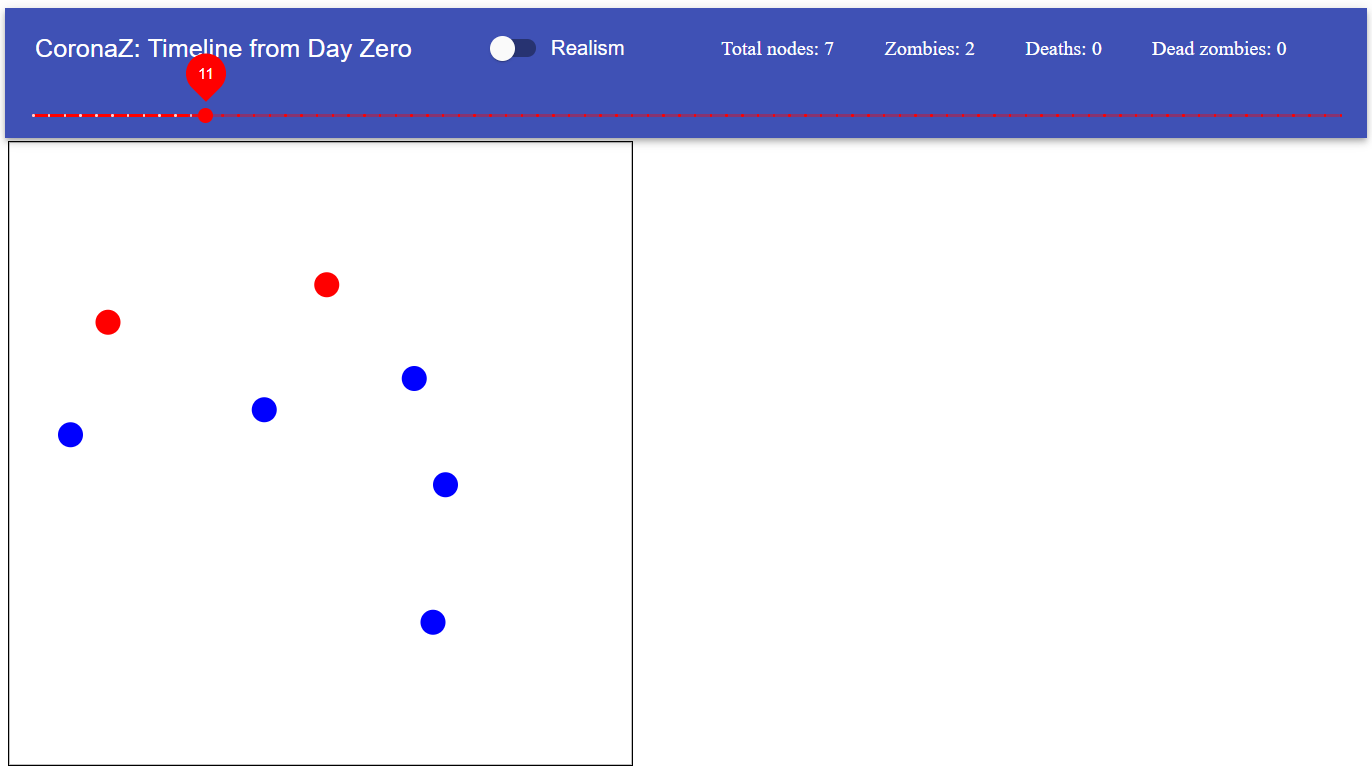}}
			\caption{CoronaZ front-end.}
			\label{fig:front-end}
		\end{figure}
	
		Safe nodes are colored in blue while infected nodes are red. Dead nodes are gray. 
		
		The statistics are the following: ``Total nodes'', ``Zombies'', ``Deaths'' and ``Dead zombies'' which respectively represent the total number of nodes, number of infected nodes, number of dead nodes and number of nodes that died infected.

\section{Docker networking}\label{sec:network}

	Since this project utilizes Docker containers, there is a docker network called \texttt{coronaZ}, defined in the \texttt{docker-compose.yml} file, which connects all the components.
	
	This network allows each node to get a new IP address and to have their own ports for broadcasting and listening.
	
	Setting the network mode in another way, for example with \texttt{network mode:host} is inadvisable due to operability issues between docker for Windows and Linux, as well as creating problems with IP and port assignment.
	
\section{Scalability and fault tolerance}\label{sec:scalability}

	We have tested the scalability and the fault tolerance of the project in the following ways:
	
	\begin{itemize}
		
		\item \textit{unexpectedly shutting down the MongoDB database}: when MongoDB fails, the user is presented with a loading \texttt{gif} in the front-end that communicates an error happened in the system.
		The DB consumer will wait to send the messages until the database itself becomes reachable again;
		
		\item \textit{unexpectedly shutting down the DB consumer}: if the DB consumer goes offline Kafka will still hold the messages in the ``\textit{coronaz}'' queue until asked for them.
		The database will not be updated until the consumer comes back up;
		
		\item \textit{unexpectedly shutting down the back-end}: as per the database, if the back-end fails, a loading \texttt{gif} will be presented to the user in the front-end.
		When it comes back up, it will serve the front-end again and the user will see the current map and simulation;
		
		\item \textit{unexpectedly shutting down the broker}: this would make nodes not able to communicate with the rest of the system, they would only send messages among themselves. 
		These messages will be stored in the nodes until the broker becomes available again.
		When the broker comes back up again the messages will contain all the contacts that have been recorded by each node, but will not contain all the movements made.
		The next message will contain the movement from the current position and the movements made in the downtime of the broker will be lost.
		The movements are not essential since the exact movements are not relevant in a contact tracing application, the most important data comes from the contacts;
		
		\item \textit{adding more nodes}: when a node is added it will start communicating with the other nodes already in the system and it will start sending messages to the broker;
		
		\item \textit{unexpectedly shutting down a node}: if a node shuts down unexpectedly, the last message with the ``\texttt{"alive": false}'' parameter will not be sent to the server.
		This would cause the front-end to keep showing the node as a fixed point on the map.
		The rest of the system will not be affected by a node shutting down.
		
	\end{itemize}

	In general for the errors that can be felt in the front-end, such as the database not responding or the back-end server not responding, the user does not really need to know what exact component went down.
	This is also for security reasons in case a \textit{malicious} user wants to find the vulnerabilities; without the specific logs it is harder for him to find the exact breaking point of the project.
	
	Also the user does not need to concern himself with the various components which make up the whole network, he perceives the system as one entity.
	
\section{Simulation}\label{sec:simulation}

	To start the project we have made an \texttt{init-project.sh} script that asks the parameters with which the simulation will take place, executes the \texttt{docker-compose} (\texttt{up} and \texttt{down}) commands and manages the number of nodes by spawning as many as the user wants.
	
	\begin{figure}[htbp]
		\centerline{\includegraphics[width=\linewidth]{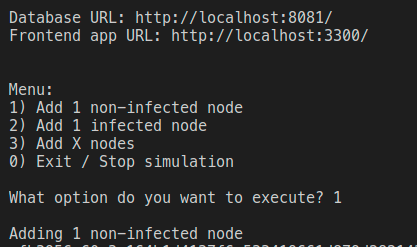}}
		\caption{CoronaZ front-end.}
		\label{fig:frontend}
	\end{figure}

	The script will ask for the run parameters and will set them on the file, otherwise will run with a set of defaults.
	The default parameters for the run are (in \textit{json} format):
	\begin{lstlisting}[language=json]
    {
        "field_width" : 100,
        "field_height" : 100,
        "scale_factor" : 5,
        "zombie_lifetime" : 120,
        "infection_radius" : 2,
        "infection_cooldown" : 15
    }
	\end{lstlisting}
	
	These parameters stand for:
	\begin{itemize}
		\item ``\textit{field\_width}'': width of the map;
		\item ``\textit{field\_height}'': height of the map;
		\item ``\textit{scale\_factor}'': scales the map and the nodes in order to be viewed better in the front-end;
		\item ``\textit{zombie\_lifetime}'': lifetime in seconds of the nodes;
		\item ``\textit{infection\_radius}'': the distance that the nodes can consider to infect other nodes;
		\item ``\textit{infection\_cooldown}'': ``\textit{cooldown}'' period in seconds in which the nodes stand in order to ``\textit{cure}''. This value is also used as an ``\textit{immunity period}'' in which the node cannot get infected.
	\end{itemize}
	
	\begin{figure}[htbp]
		\centerline{\includegraphics[width=\linewidth]{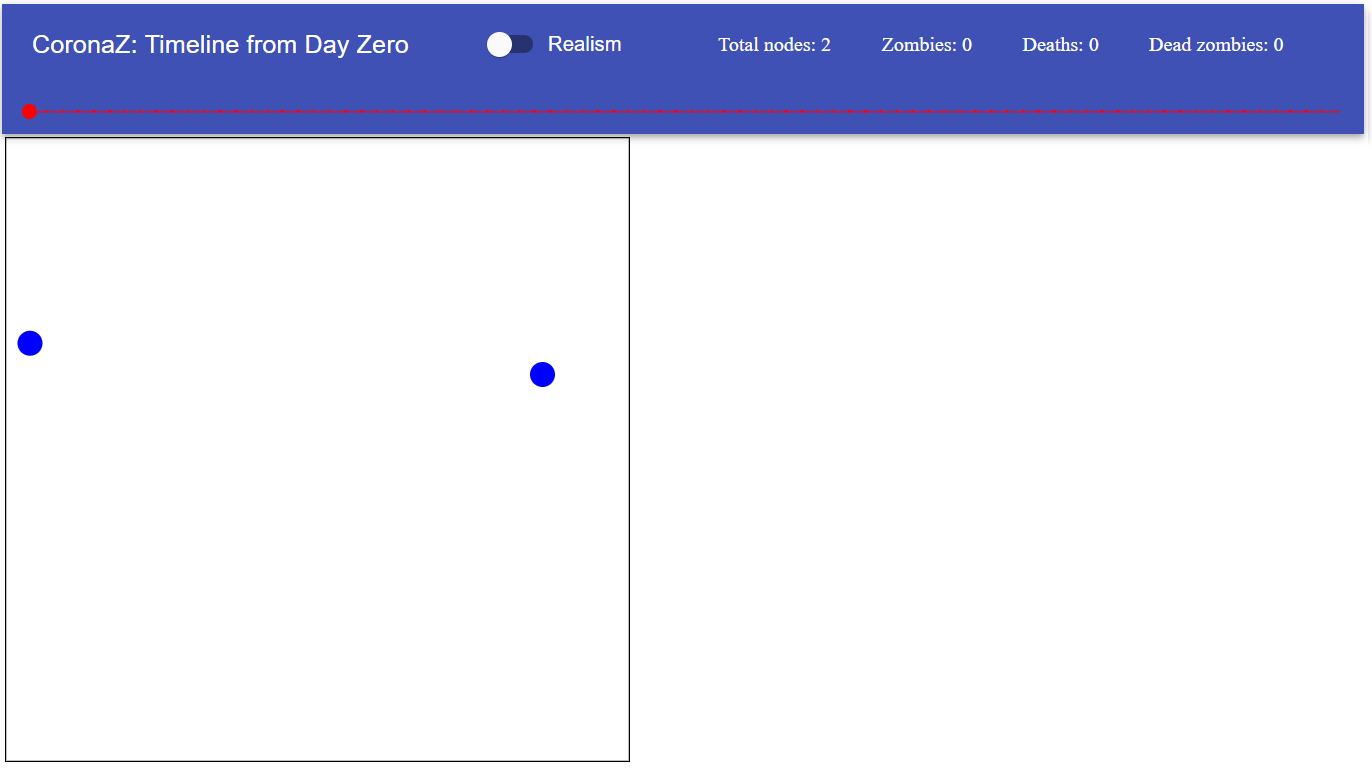}}
		\caption{CoronaZ simulation at start.}
		\label{fig:sim_1}
	\end{figure}
	\begin{figure}[htbp]
		\centerline{\includegraphics[width=\linewidth]{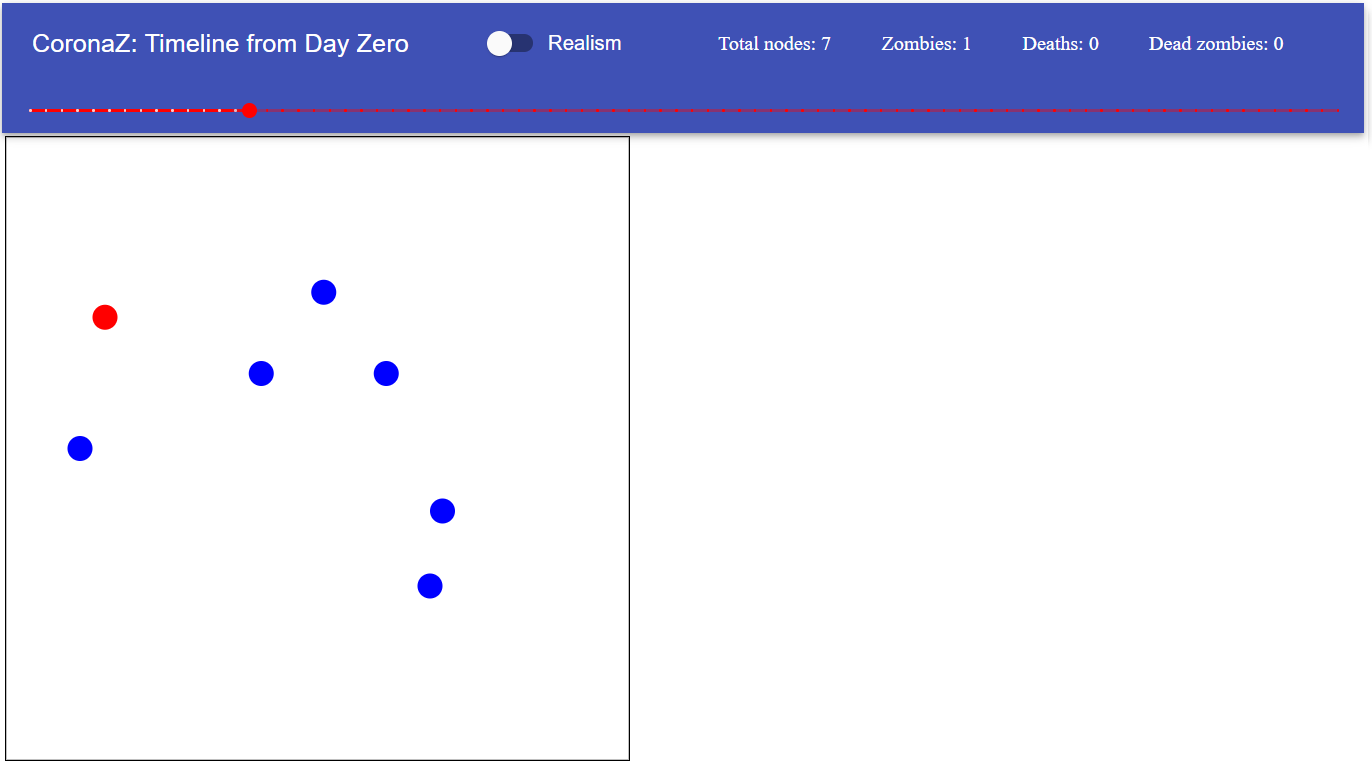}}
		\caption{CoronaZ simulation when first node is cured.}
		\label{fig:sim_2}
	\end{figure}
	\begin{figure}[htbp]
		\centerline{\includegraphics[width=\linewidth]{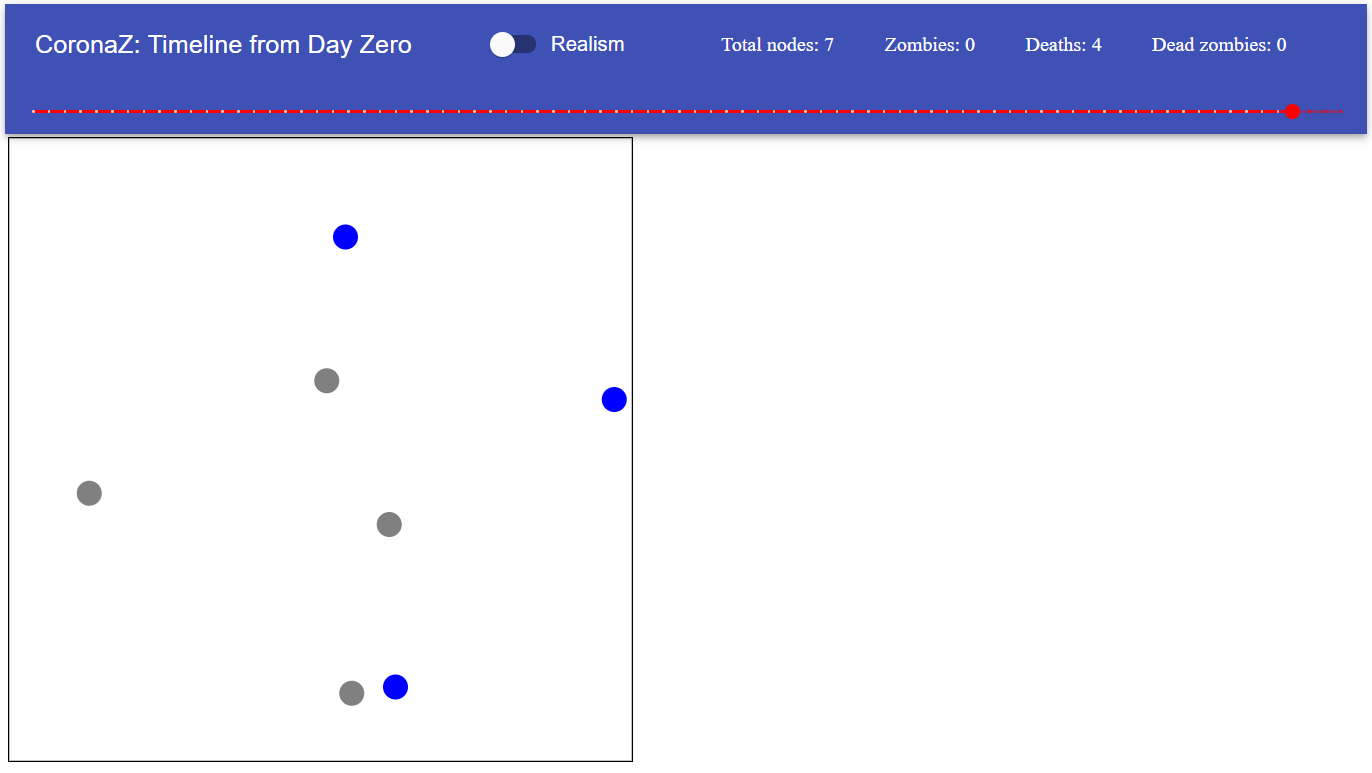}}
		\caption{CoronaZ simulation when nodes die (from natural causes).}
		\label{fig:sim_3}
	\end{figure}

\section{Performance evaluation}\label{sec:performance}

	A ``\texttt{simulation.pcapng}'' file is present in the essay folder and contains part of a simulation for the system system running.

	From this file, various message lengths can be seen, for example: the length in bytes of a message sent in broadcast by each node is \textit{148B} (just the payload, \textit{189B} with headers).
	For the messages sent to the the server by the nodes, the minimum size is \textit{148B} (\textit{189B} with headers), the rest depends on the number of contacts the node had.
	
	For a simulation with twenty nodes, each of them sends one message in broadcast a second, which means 189B.
	In the ``\textit{worst case scenario}'' where all the nodes are in infecting range, the lenght of the message sent to the server would be \textit{2.49KB}.
	
	So just one node sends $189~B + 2.49~KB = 2.68~KB$ each second, this would mean that for a simulation of $120$ seconds, the total amount of data sent by the nodes is $2.68~KB*120*20 = 321.6~KB*20 = 6.43~MB$.
	
	Latency does not represent an obstacle in our system since each node stores the messages that need to be sent to the server and the data handled is not time sensitive as per other \textit{real-time} applications.
	
	Network performances for the nodes are discussed in \ref{subsec:nodes_improvements}.

\section{Future work}\label{sec:future_work}

	Given the nature of the project there won't be future releases but in this section we want to talk about what can be done to improve the project.
	
	There are various improvements that can be done to make CoronaZ a more interesting and stable simulator, we will tackle them based on the areas defined in \ref{sec:architecture}.
	
	\subsection{Nodes}\label{subsec:nodes_improvements}
	
		Nodes logic can be improved adding a more intelligent behaviour.
		This would result in better movements on the map and, possibly, avoiding infected nodes.
		
		The throughput of the network could be improved by shortening the messages and choosing a different message format (like \textit{differential updates}).
		This would allow not only to exchange less bytes between the nodes but especially between the nodes and the Kafka server.
	
	\subsection{MQTT broker}
		
		In order to allow for a better network performance, scaling the broker would be the best option.
		A dynamic scale rule could be created and the server replicas can be hidden behind a proxy.
		
		This would not only help improving the performance of the network but also by adding fault tolerance in case one of the replicas fail.
		
	\subsection{DB consumer}
		
		As per the broker server, scaling the DB consumer node and making each node read from a separate queue or a separate server would improve network performance and fault tolerance.
	
	\subsection{Database}
	
		Currently the database only holds one simulation, but it can be expanded to hold multiple ones, preferably as different MongoDB collections.
		
		As MongoDB is very configurable, one could also tweak factors such as replication.
	
	\subsection{Back-end}
		
		Currently the back-end only supports one end-point. 
		An important improvement would be to allow the front-end to make intelligent queries, for example only sending the newest states, instead of all of them, to reduce bandwidth consumption. 
		
		Additionally, more complex functions such as selecting a specific simulation or user profiles can be implemented.
	
	\subsection{Front-end}
	
		The front-end is feature complete in terms of our project scope.
		A further improvement could be an option to select a specific simulation or having a user profile.
	
\section{Conclusions}\label{sec:conclusions}

	This project represents a simulation of a contact tracing application where each node that spawns in the network communicates its position to the others in range and will send the data collected to a central server with the \textit{publish-subscribe} pattern.
	
	The running system implements the basic goals and functionalities requested in the given programming task sheet.
	As explained in \ref{sec:scalability} it supports scalability and fault tolerance, while the end user of the system can decide how many nodes to add, thus scaling the system \textit{horizontally}.
	In \ref{sec:performance} there is the analysis of how the system performs and how much data is sent in the network during the simulation.
	Improvements, both in the system's architecture and it's performance are discussed in \ref{sec:future_work}.
	This project has these three characteristics: \textit{naming and node discovery}, \textit{synchronization and consistency} and \textit{fault tolerance}.


\begin{thebibliography}{00}
	
	\bibitem{coronaz_repo}
		\textit{CoronaZ repository:}\\
		\url{https://github.com/cipz/CoronaZ/}
	
	\bibitem{uuid}
		\textit{\texttt{uuid} — UUID objects according to RFC 4122:}\\
		\url{https://docs.python.org/3.8/library/uuid.html}
		
\end{thebibliography}
\end{document}